\documentstyle[11pt]{l-aa}

\begin{document}
\thesaurus{03 (11.09.4; 11.09.1; 09.04.1; 09.13.2)}
\title{FIR and C$^{+}$ emissions of spiral galaxies disks.}
\subtitle{The example of NGC 6946.}
\author{S. Sauty\inst{1}, M. Gerin \inst{2,1}, F. Casoli\inst{1}}
\institute{DEMIRM, Observatoire de Paris, 61 Av. de 
l'Observatoire, 75014 Paris, France; and URA 336 du CNRS 
\and  Radioastronomie Millim\'etrique, ENS, 24 Rue Lhomond, 75231 
Paris Cedex 05, France; and URA 336 du CNRS}

\offprints{M. Gerin}
\date{Received 15 October 1997; accepted 25 May 1998}
\maketitle

\begin{abstract}

We present numerical simulations of radiative transfer in the spiral galaxy 
NGC 6946.  The interstellar medium is represented as
a two phases medium, with molecular clouds and a smooth
diffuse phase. The molecular gas  distribution is calculated in a 
self-consistent way from
the distribution of an ensemble of molecular clouds evolving
in the gravitational potential of NGC 6946.
We simulate star formation by creating OB associations in
molecular clouds.
 The transfer of UV radiation is calculated in the clumpy 
interstellar medium, to determine the local UV illumination 
of molecular clouds.
 We compute the emergent intensity in the UV continuum 
(912 -- 2000 \AA),
in the H$\alpha$ and  C$^{+}$ $^2P_{3/2} - ^2P_{1/2}$ lines 
as well as in the continuum at far infrared wavelengths, 60, 
100 \& 200 $\mu$m.

It is possible to obtain a consistent picture of this galaxy with
a global star formation rate of 4 M$_{\odot}$yr$^{-1}$ 
(for stars with masses in the range 2-60 M$_{\odot}$) 
occuring
mostly in the spiral arms.
 The close spatial association
of massive stars and molecular clouds has a profound impact on the
transfer of UV radiation in the galactic disk and on the 
dust emission. The median distance travelled by 
UV photons is about 120 pc. However, when they have escaped 
from the vicinity
of their parent OB associations, UV photons may travel
quite far in the disk, up to 1 kpc. The UV opacity of the model
spiral galaxy disk, observed face-on, is 0.8 at 1000 \ \AA \ and 0.7 at 2000 \AA.

For radii less than 4 kpc, the C$^{+}$ 158 $\mu m$ line is
mostly produced in photodissociation regions at the
surfaces of molecular clouds. The C$^{+}$ emission from
diffuse atomic gas accounts for about 20\% of the total. It becomes
significant at large distance from the nucleus ($r \geq 4$ kpc).
Molecular clouds and diffuse atomic gas have almost equal contributions to
the total far infrared emission from 60 to 200 $\mu m$. As a whole, 72\% of the
60 -- 200 $\mu m$ FIR emission can be attributed to dust grains
heated by the UV radiation of massive stars and 
 28\% by the radiation field of the old stellar population.\\


\end{abstract}

\section{Introduction}

 Because the peak of the thermal dust emission occurs in the far-infrared 
(60-200 $\mu$m)
 for Galactic interstellar clouds, this wavelength domain  is recognized as a
very important `window` to study the interstellar medium,
both in the Milky Way and in external galaxies. However, due to the lack of
spatial resolution provided by  telescopes operating at these wavelengths,
the interpretation of the signals measured in external galaxies
is still a subject of controversy. The debate is focussed on the origins of the
 heating of dust grains and of the 
$^2P_{3/2}-^2P_{1/2}$  C$^+$ line at 158 $\mu$m. 
This fine structure line 
  of ionized carbon is one of the main cooling lines of neutral atomic gas 
(Wolfire et al. \cite{wo}), but  it
 is also seen in HII regions, with different excitation mechanisms in 
these two cases. Other sources of C$^+$ emission
are the dense photodissociation
regions (PDRs) where intense UV radiation from massive stars impinges on
 molecular clouds. In these regions,
the photodissociation of carbon monoxide creates 
a layer of ionized carbon at the edge of the molecular cloud, which
is a source of intense C$^+$ emission.

The far infrared emission is produced by dust grains heated by star light but 
the relative contribution of young  massive stars on one hand, and of the
bulk of the stellar population on the other hand, are still debated
 (see e.g. Walterbos \& Greenawalt \cite{wg}, Devereux \& Young
\cite{dy}, Cox \& Mezger  \cite{cm}, Thronson et al. \cite{th}, 
Rice et  al. \cite{ri}).
Using low resolution data from IRAS,  IR colors have been used to assess
the respective roles of star forming regions and cirrus in external
galaxies at a global scale (e.g. Calzetti et al. \cite {calzetti}).  
It is difficult to find spiral galaxies dominated by a single population
of heating sources, since the relative contributions of massive stars and of
the disk population vary from galaxy to galaxy, and probably from place
 to place in a given galaxy. A main problem of these studies is the lack of
spatial resolution at far infrared wavelengths.  
 Even when it is possible
to use high spatial resolution data at millimeter and submillimeter
wavelengths, the gap in spatial resolution
from the local interstellar clouds to  external galaxies
is huge: for nearby galaxies located at distances $D$
of  a few Mpc, a one arcmin beam encompasses $0.3 (D/1 {\rm Mpc}) {\rm kpc}$, 
much larger
 than the size of a single molecular complex in the Galaxy, whereas 
structures down to 0.01 pc are commonly observed in local
clouds (Falgarone et al. \cite {fpp}) and in the  Magellanic Clouds 
(Rubio et al. \cite {rl}).

Numerical simulations provide a way out of this problem: it is possible
to reproduce the observed molecular gas distribution
 of nearby spiral galaxies with 
numerical simulations using observed data as
input parameters. For example  Garcia-Burillo et al. (\cite {garcia}) were
able to fit the spatial distribution and kinematics of the
molecular gas in M 51 using the cloud collision code developped
by Combes \& Gerin (\cite {cg}). 
Though  this code has a spatial resolution of a few hundred parsecs (the
cell size for the large scale dynamics),
it is possible to include "micro-physics" at the parsec scale inside each cell.
We have taken this   approach and implemented star formation 
in this code  to study the far infrared and C$^+$  emissions in 
the spiral galaxy NGC 6946. The next section summarizes the current data on 
NGC~6946. We describe the model in Sect. 3 and present  the results for 
NGC~6946  in Sect. 4. The implications of this work  for the interpretation 
of the C$^+$ and FIR emission of spiral galaxies  are discussed in Sect. 5.

\section {NGC 6946}
NGC~6946 is a nearby Scd galaxy with a low inclination, $i = 34^\circ$
 (Consid\`ere \& Athanassoula \cite {ca}).  At the adopted distance of 5 Mpc,
 1 arcmin on the sky corresponds to a linear size of 1.5 kpc.
The main characteristics  of NGC~6946 are summarized in Table 1. 
 It shows a weak bar (Martin 1995) and a 
prominent spiral structure with both an $m=2$ and an 
$m=4$ pattern (Consid\`ere \& Athanassoula \cite {ca}). With this
open spiral structure, the arm and interarm regions are  
resolved at moderate spatial resolution (30''). The gas  distribution is
well known since NGC~6946 has been mapped at high spatial resolution in
HI (Boulanger \& Viallefond \cite {bv}), and CO (Casoli et al. \cite {cf},
Clausset et al. \cite {ccf}). Figure 1 presents a CO(2-1) map
 obtained at the IRAM 30m telescope (Clausset et al. \cite {ccf}), 
and Fig. 2 a  V-I color map  obtained  at the Observatoire
de Haute-Provence (OHP) (P. Boiss\'e, private communication).
 NGC 6946 is marginally resolved in the far infrared
maps obtained by IRAS and the KAO  (Engargiola \cite {engargiola}). It also 
belongs to the normal galaxy sample studied by ISO and as such has
been extensively mapped in the mid and far infrared (Malhotra et 
al. \cite {malhotra}, Helou et al. \cite {helou}, Lu et al. \cite {lu}, 
Tuffs et al. \cite {tuffs}). NGC~6946 is forming stars actively, as
revealed by its  bright H$\alpha$ emission (Bonnarel et al. 1988a, 1988b,
 Kennicutt \cite {kennicutt})  and the numerous supernovae 
(Li \& Li \cite {li}). Madden et al. (\cite {sm}) have mapped the C$^{+}$
158 $\mu m$ emission at 45'' resolution  with the KAO. Bright C$^{+}$ 
emission  is found in the nucleus and in the disk.
At this scale,  the C$^{+}$ emission is correlated with the CO emission 
in the spiral arms and with the  HI emission in the outer disk.

\section{A model of NGC~6946}
\subsection{Representation of the interstellar medium.}
 We have chosen to include two phases for the interstellar medium :\\
i) a dense molecular  phase composed of spherical molecular clouds with
masses ranging from 10$^3$ to 10$^6$ M$_\odot$ and mean density
50 H$_{2}$ cm$^{-3}$.  This mean density is used to determine cloud sizes 
from their masses. The half power width of the molecular layer, defined as 
$<z^{2}>^{1/2}$, is equal to 65 pc.  The global volume filling factor of this
 molecular phase in the disk, $\Phi_{V}$ (H$_2$), amounts to 2.5\% .
The clumpy structure of molecular clouds  is taken into account by using a 
larger density,  $5 \times 10^{3}$ H$_{2}$ cm$^{-3}$, to compute the 
C$^{+}$ emission. The C$^{+}$ emission is not sensitive to the gas density 
when it is larger than $\sim$ 10$^3$  cm$^{-3}$ (see Tielens \& Hollenbach  
\cite {th85} and Sect. 3.6). Indeed,  C$^{+}$ observations of nearby 
molecular clouds show that C$^+$  emission  is detected  over the whole 
projected surface of  molecular  clouds (Jaffe et al. \cite {jaffe}).  
The total mass of dense molecular gas  in the model is set to match the 
value deduced from CO(1-0) observations with a conversion factor of 
$2.3 \times 10^{20}$ H$_{2}$cm$^{-2}$
 (K km$s^{-1}$)$^{-1}$, $2.0 \times 10^9$ M$_\odot$ (Table 1).\\
ii)  a neutral diffuse phase, with a nearly constant HI density. 
We have adjusted the value of the  mean HI density using the data from 
Boulanger \& Viallefond (\cite {bv}): it varies slowly from 0.8 H cm$^{-3}$ 
at the center to 2 H cm$^{-3}$ in the outer disk ($r \sim$ 6 kpc). 
 We have assumed that the HI disk  has a vertical thickness of 
360 pc = $2 H_z$, and that the density is uniform from $-H_z$ to $+H_z$.
The velocity dispersion of the diffuse gas is set to 10 kms$^{-1}$. 
The total mass of diffuse neutral gas inside a radius of 6 arcmin
is taken from the HI observations to be $2.6 \times  10^9$ M$_\odot$. 

 In addition to these two phases, we take into account the
formation of  HII regions around OB associations (see below).
For the numerical calculations,  we represent the galaxy disk as a large 
3D grid. The total grid length $L$ and the cell size $r_{cell}$ are 
chosen to match the following constraints: \\
-- the grid size has to be at least as large as the optical disk, which has a
diameter of 12.5 kpc, in order to be able to follow the path of photons in 
the external  regions of the galaxy, and the heating of the neutral gas at 
large distance from the nucleus. \\
-- the spatial resolution, $r_{cell}$, should be good enough to describe 
the clumpy structure  of the medium: in particular the smallest clouds
in the ensemble should occupy at least one cell. With the
adopted parameters for the clouds, the radius of
the smallest clouds with a mass of 10$^3$ M$_\odot$ is 4.5 pc. 
This number is only an estimate of the actual size of a given cloud. 
It is possible to use an alternative method based on the scaling 
relations for molecular clouds, the so-called Larson's laws. 
Using the mass-radius relation
$M ({\rm M_\odot}) = 100 R^2 ({\rm pc^{2}})$ (Falgarone et al. \cite {fpp}),
 the radius of a 10$^3$ M$_\odot$ cloud is then 3 pc.
Another constraint on the spatial resolution is provided by the size
of OB associations, which is generally larger than 30 pc (Garmany \cite 
{garmany}) and their distance from molecular clouds.
Leisawitz (\cite {leisawitz}) has estimated  the mean distance between OB 
associations and their parent cloud to be about 50 parsecs.

We have chosen as the best compromise to represent the galaxy with a
$1024 \times 1024 \times 24$ cell structure. The galaxy size is then 
$L = 2 \times R_{max}$ = 12.5 kpc and the resolution $r_{cell}$ = 12.2 pc.
The cells are filled with neutral atomic gas, or with molecular gas at
 the molecular cloud positions, an 10$^{6}$ M$_{\odot}$ molecular cloud 
 then occupies $3\times 3 \times 3$ cells.
Finally, we include the ionized gas in the Str\"omgren spheres centered on 
each OB association, and replace  the molecular and atomic  gas with
 ionized gas whenever required. The
radii of the Str\"omgren spheres are calculated assuming 
 classical ionization in HI-bounded HII regions (Miller \& Cox \cite {mc}).
Because of the coarse spatial resolution of the dynamical
code (200 pc), we do not attempt to reproduce the interstellar medium in
 the central region  of NGC 6946 ($r < 500$ pc).

\subsection{Spatial distribution of molecular clouds}

This section of the code uses the cloud-cloud collision code described
by Combes \& Gerin (\cite {cg}), and used by Garcia-Burillo et al. 
(\cite {garcia}) and Gerin et al. (\cite {gerin}) to model the gas dynamics 
in nearby galaxies. The molecular clouds move in the gravitational potential 
of the galaxy, they  grow through cloud-cloud collisions, which are sticky 
processes, and are disrupted by  simulated SN events. The molecular gas is 
immediately recycled into small molecular clouds.  The gravitational potential
 is deduced from an  R band image of NGC 6946 (Viallefond  \& Bonnarel, 
private communication, Bonnarel et al. \cite {bonnarel}), with foreground 
stars removed. The image has been rotated to put the major axis vertical, 
and deprojected to face-on, using as  projection parameters PA = 69$^\circ$
and i = 34 $^\circ$. The image has then been binned to  $256 \times 256$ 
pixels.  This R band image covers $8.5' \times 8.5'$ on the sky, corresponding
to a radius of 6.24 kpc at the assumed distance of NGC 6946, 5 Mpc. The actual 
spatial resolution amounts to a few times the cell size, about 200 pc. 
The following step is to build a good gravitational potential from this image. 
As in Garcia-Burillo et al. (\cite {garcia}), the calculation is done in two 
steps. The  axisymmetric part of the potential is obtained by assuming a 
constant mass to light ratio,  and adjusting this constant to reproduce
the observed CO and HI rotation curves. The perturbations due to
the bar and spiral arms are included in the non axisymmetric part of the
potential. The last parameter to adjust is $\Omega_{p}$, the pattern speed of
 the density wave (bar + spiral arms). The gas distribution and velocity field
 are very sensitive to $\Omega_p$.  We find that $\Omega_{p}$= 42 
kms$^{-1}$ kpc$^{-1}$ gives the best results.  The corotation is found at a 
 radius of  3.5 kpc, close to the end of the bar,
 and the OLR lies outside of the disk. Figure 3a presents the deprojected 
R-band image. Figures 3b and 3c present an example of the obtained molecular 
cloud distribution. The gas clouds are concentrated in the spiral arms, with
few molecular clouds at radii larger than 4 kpc. The model rotation curve 
is shown in Fig. 4, together with the angular frequency $\Omega (r)$. 
We have not attempted to model the compression of diffuse gas
in the spiral structure, and have kept an axisymmetric distribution
of diffuse atomic gas.

\subsection{Stellar population}

Once the gas is distributed over the galactic disk, 
we create OB associations, with star masses ranging from 10 to 60 M$_{\odot}$ 
(stellar types from B2 to O5). We have chosen not to include  stars less 
massive than 10 M$_\odot$ because  their lifetime becomes a significant 
fraction of the rotation period. Shortward of 2000 \AA, the UV radiation is 
mostly produced by OB stars and the contribution by late B type stars is at 
most 20\% (Walterbos \& Greenawalt \cite{wg}, Mathis et al. \cite{mmp}).
 The choice of the upper mass cutoff at 60 M$_\odot$ is motivated by the work 
of Heydari-Malayeri \& Beuzit (\cite {hmb}) who have shown that suspected 
very massive stars ($\sim$ 100 M$_{\odot}$) are actually clusters of less
 massive stars. Also, a very short time step is 
required to sample adequately  the lifetime of these 
very massive stars.

  We allow only clouds more massive than $4 \times 10^4$ M$_\odot$ 
to form massive stars. In the cells satisfying this criterion, 
the OB associations  are born at the outer edge of the molecular clouds. 
We have tested different star formation laws, so as to match as
well as possible the H$\alpha$ radial profiles. The best fits are 
obtained with a star formation  rate (SFR)  depending on the angular 
frequency and the local gas surface density as proposed by Wyse \& Silk 
(\cite {ws}) :
 $$ SFR( \vec{r}) =  \epsilon \Omega(r)\sigma_{gas}(\vec{r})$$ 
with $\epsilon$ the star formation efficiency, $\Omega (r)$ the 
angular frequency at radius $r$ and $\sigma_{gas} (\vec{r})$ the 
HI+H$_2$  surface density at position  $\vec{r}$ in the disk.
 We find that $\epsilon = 5 \%$ provides the best match  to the 
H$\alpha$ radial profile by Kennicutt (\cite {kennicutt}). 

The stellar mass distribution inside an OB association is drawn from the
 Initial Mass  Function (IMF). We have chosen an index close
to the Salpeter IMF, as suggested by measurements in Galactic and 
extragalactic OB associations (Massey et al. \cite {massey}) :
  $\frac{dN(M)}{dM} = M^{-2.3}$.

The stellar associations are born with a mean velocity relative
to their parent cloud chosen from a Gaussian of mean value 10 kms$^{-1}$ 
and FWHM $\sim$ 10 kms$ ^{-1}$. These figures are in good agreement with 
observed data by Leisawitz et al. (\cite {lbt}) in their study of the 
relation of star clusters with molecular clouds. With this value, the
average  motion of OB stars relative to their parent cloud is  10 pc in
 10$^{6}$ years.  Finally, stars die after a time equal to their  Main 
Sequence lifetimes   from G\"usten \& Mezger (\cite {gm}). Effective 
temperatures, stellar luminosities and radii, averaged over the main-sequence 
lifetime, are taken from Cox et al. (\cite {cox}). Because we are only 
interested in broad 
band fluxes and colors, we have calculated  all stellar fluxes, from UV 
to B bands, in the black-body approximation. Lyman continuum radiation 
production rates, averaged on the Main Sequence lifetime, are also taken from 
 Cox et al. (\cite {cox}) and G\"usten \& Mezger (\cite {gm}) to
 have a coherent set of parameters. We use the parametrization as a function 
of the stellar mass, $M$: 
$$ log_{10} (\frac{N_{Lyc}}{s^{-1}} ) = 38.3 + 8.16 \times log_{10}(M) $$  
$$ -0.24 \times log_{10}^{2}(M) - 0.41 \times log_{10}^{3}(M) .$$ 
to compute the thermal radio continuum flux, H$\alpha$ luminosity and 
Str\"omgren sphere diameters.

We also include the effects of the radiation of massive stars on neutral gas,
 and allow molecular clouds to be partially eroded and ionized by the 
radiation of nearby OB associations, if they overlap with the Str\"omgren 
sphere of an OB association. We compute the radius of this sphere, 
$R_{HII}$, assuming that all stars are located at the same position and 
have a global production rate of Lyman continuum photons  $N_{Lyc}$, 
equal to the sum of the contribution of the individual stars, and for case B 
recombination in a diffuse medium of density $n_H$: 
$R_{HII} = (\frac {3}{4} \frac {N_{Lyc}}{\pi \alpha_B n_H^2})^{1/3}$. 
In this formula, $\alpha_{B}$ is the hydrogen recombination coefficient 
(Osterbrock \cite {osterbrock}). The mean radius for R$_{HII}$ is 35 pc.

To constrain the population of massive stars, we calculate different
stellar outputs, namely we perform a detailed calculation of the UV radiation 
field at 912-2000 \AA, and also compute the global U flux and U radial 
profile, as well as the  thermal radio continuum and H$\alpha$ emission.

$\bullet$ 6 cm radio-continuum.\\
According to Mezger (\cite {mezger}) and Turner \& Ho (\cite {turner}), 
the radio continuum emission of HII regions for case B recombination, 
at an electronic temperature  $T_{e}$ of $10^4$ K  and with 45 \% of the 
ionizing photons being converted into H$\alpha$ photons, is directly 
related to N$_{Lyc}$ by: 
$$ \frac{N_{Lyc}}{s^{-1}} = 1.1 \times 10^{50} (\frac{S_{6 cm}}{mJy})
  (\frac{D}{Mpc})^{2}. $$
This formula does not include any correction for  dust absorption of the 
ionizing radiation within the HII region. Current estimations are that 
nearly 50\% of the Lyman continuum may be absorbed by dust. 

$\bullet$ H$\alpha$.\\
The relationship between the H$\alpha$ luminosity  and the production rate of 
Lyman continuum photons, using the same assumptions as above, can be deduced 
from Mezger (\cite {mezger}) and Peimbert et al. (\cite {peimbert}):
$$ \frac{L_{H\alpha}}{L_{\odot}}=  3.55 \times 10^{-46} (\frac{N_{Lyc}}
{ s^{-1}}) .$$
The correction for the extinction may amount to about 1 magnitude at 
H$\alpha$ but it is highly uncertain (Mc Kee \& Williams \cite {mkw},
 van der Hulst et al. \cite {vdh}). Because of the uncertainties involved
in this correction, we have preferred not to do it. This allows us
to check that the energy is conserved with a good accuracy in the
simulation.

 The formed OB associations have typical  production rates of  Lyman 
continuum photons, $N_{Lyc}$, in the range 10$^{45}$-10$^{51}$ s$^{-1}$.
 The cumulative Lyman continuum  luminosity function of the 
population is shown in Fig. 5.  Its shape is similar  to the distribution 
for Galactic HII regions (McKee \& Williams \cite {mkw}). There is a small 
excess around $N_{Lyc}$ $\geq$ 10$^{49.5}$ s$^{-1}$, dominated by the most 
massive stars in the younger associations, and a deficiency of associations 
with $N_{Lyc}$ $\geq$ 10$^{50.5}$ s$^{-1}$. As a whole, the match of the 
two distributions is very good. The distribution of the intrinsic UV 
luminosity of OB associations extends over four orders of magnitude from 
10$^{4}$ to 10$^{7}$ L$_{\odot}$, i.e. from small associations with about 
20 stars and a total stellar mass of 600 M$_{\odot}$, up to large associations
gathering 200 OB stars, including a few 50-60 M$_{\odot}$ stars,
 and having a total stellar mass of 6000 M$_{\odot}$.

\subsection{Dust properties}

We use the average Galactic extinction curve from Fitzpatrick \& Massa
 (\cite {fm})  at UV wavelengths. For visible and near infrared wavelengths,
we use the work by Seaton (\cite {seaton}). This curve is probably 
valid on a large scale in the diffuse medium of NGC~6946 which has a 
similar metallicity to the Milky Way. We assume a constant metallicity and 
gas to dust ratio in the disk, and use the average value for the Milky Way :
 $\frac{NH}{E(B-V)} = 5.8 \times  10^{21}$  H cm$^{-2}$ mag$^{-1}$ 
and A$_{V}$ = 3.1 $E_{B-V}$ (Bohlin et al. \cite {bohlin}). 
The extinction through a cell filled with molecular gas,
 with the adopted spatial resolution, is 2.5 magnitudes.
The dust  properties have been summarized by Bruzual et al. (\cite {bruzual}) 
and Witt \& Gordon (\cite {wg96a}). Apart from the enhanced
absorption in the 2175 \AA \ bump, the dust albedo
is fairly constant at UV and visible wavelengths at $\omega \sim 0.55$.
We include the coherent scattering of UV light by dust grains.
 The anisotropy is described  using the Henyey-Greenstein (1941) function, 
where the anisotropy parameter is defined as $ g = < cos \theta >$ and 
$\theta$ is the scattering angle.

\subsection{Radiative transfer}
The observed properties of external galaxies depend on the
propagation of the stellar radiation in the interstellar medium. To determine 
the local radiation field, we follow the propagation of UV photons 
(912-2000 \AA)  emitted from the OB stars in the  two phases medium.
In each cell, we compute a local radiation field. We define the local 
radiation field in the UV, $\chi_{UV}$, relative to 
the mean radiation field  in the UV at the Solar radius, $\chi_{0}$, 
established by Mathis et al. (\cite {mmp}), so that 
$\chi_{UV} = \frac {4 \pi J} {\chi_0}$. In the Galaxy and at
the solar radius, the InterStellar  Radiation Field, ISRF in the UV, 
has been defined as $\chi_{0}= 4 \pi J = 4 \pi \int_{912 \AA}^{2000 \AA}
 J_{\lambda} d\lambda $ = $1.84  \times 10^{-3}$ erg cm$^{-2}$s$^{-1}$ 
and through visible/IR bands as G$_{0}$ = $4 \pi \int_{912 \AA}^{2 \mu m} 
J_{\lambda} d\lambda $ = $2  \times 10^{-2}$ erg cm$^{-2}$s$^{-1}$, 
where J$_{\lambda}$ is the specific intensity of the
radiation field averaged over 4$\pi$ sr (Mathis et al. \cite {mmp}).

On their path, photons can be scattered and/or absorbed by dust in both 
neutral phases. We have chosen not to follow individual photons which 
would have been time consuming, but to use instead  pseudo-photons 
representing a collection of $N$ photons. We are then able to probe a 
larger area  with a lower  number of photons. We sample  the 
912-2000  \AA \ interval with 20 bins of constant wavelength width 
$\Delta \lambda$, and launch n$_{\gamma}$ pseudo-photons per association 
per wavelength bin.  The longward limit has been set to 2000\ \AA, to avoid 
having to take into acount the  contribution from star types later than B 
to the ISRF (Walterbos \& Greenawalt, \cite  {wg}).

The pseudo-photons emitted by the OB association number $i$
 carry a fraction $f_{\lambda}^{i}$ $d \lambda$ 
of the luminosity $L_{\lambda}^{i}$ radiated by this OB association. 
They travel from the center of a cell (size $r_{cell}$)  to an adjacent  
one, and the energy absorbed by the interstellar medium, when non-zero, 
is left on the common face of these 2 cells. The absorbed energy is 
reprocessed in the far-infrared. The pseudo-photon  energy along its path 
from association number $i$ can be written as : 
 $$ f_{\lambda}^{i} d\lambda  =  \alpha^{i} \frac{L_{\lambda}^{i}}
{n_{\gamma} } $$
In this formula, $\alpha^{i} = \prod_j \alpha^{j}(x,y,z)$ is the product of 
the probabilities for non absorption in each cell along the travel from 
the OB association number $i$ to the cell position $(x,y,z)$, or 
with $\omega(\lambda)$ the dust albedo and $\tau_{\lambda}$ the 
total cell opacity including scattering :

$\alpha^{j}  = e^{- (1 - \omega (\lambda)) \tau_{\lambda}}$ \\
in the diffuse phase, and

  $\alpha^{j} = \omega(\lambda)$\\
 in the H$_{2}$ phase.

When leaving an OB association, the  direction for each pseudo-photon is 
uniformly chosen on the unit sphere. 
These pseudo-photons travel in the two-phase interstellar medium. 
There are different possibilities when reaching a new cell :\\
i) The gas in the cell is diffuse  and atomic. Then the pseudo-photon can 
either :\\
-- be absorbed partially in the cell. A fraction of the luminosity is left,
 and the pseudo-photon continues in the same direction  with a lower 
luminosity.\\
-- be scattered.\\ 
ii) The cell is filled with molecular gas. \\
 The pseudo-photon is partly absorbed, and partly backward scattered,
the ratio between the energy left in the cell and the total energy of the
pseudo-photon depends on the albedo as $1 - \omega(\lambda)$. We have chosen
backward scattering in that case to prevent the pseudo-photon
from interacting with the same molecular cloud several times.\\
After a scattering event, a scattering angle $\theta$ is chosen according 
to the anisotropy function, and the azimuthal angle $\phi$ is uniformly 
chosen in  the interval $[-\pi, \pi]$.  Then the new direction is
 easily deduced from the previous one  (Witt \cite {witt2}).

Since we cannot store all the pseudo-photons' incident directions, and in order
to define an isotropic radiation field in each cell, 
we assume that the equivalent surface of the cell is $6r_{cell}^{2}$. 
Then, the UV radiation field in the cell $(x,y,z)$ due to the contributions 
of all pseudo-photons travelling through this cell, can
be expressed relative to the Galactic ISRF $\chi_0$  as : 
  $$\frac{\chi (x,y,z)}{\chi_{0}}  =  \sum_{i} \sum_{\delta \lambda}
 \frac{f_{\lambda}^{i} d\lambda}{\chi_{0} 6 r_{cell}^{2}} $$
Because of the coarse angular resolution and the low number of pseudo-photons 
leaving each OB association, the resultant map of the radiation field 
presents strong fluctuations. We have chosen to smooth the map of the 
radiation field by averaging the data in nearby cells, typically 3x3 cells.
Furthermore, some cells are never visited by UV photons, for example in the 
interarm region or in the outer disk ($r \geq 5$ kpc).
In that case,  we use as incident radiation field,  the Galactic ISRF 
(longward of 2000 \AA ), scaled by the local surface brightness in 
the R band image to take into account the radial variation of the
 radiation field from  the old stellar population between the central regions
($r \leq 1$ kpc) and the outer disk ($r\geq 5$ kpc). The reference value is 
given in units of $4 \pi \int_{2000 \AA}^{2 \mu m} J_{\lambda} d\lambda $ = 
G$_{0}$ - $\chi_{0}$ = $ 1.82 \times 10^{-2}$ erg cm$^{-2}$s$^{-1}$. The 
position for the reference value  has been chosen at the edge of a spiral arm,
at a distance $r = 4$ kpc from the center, where the UV radiation 
field $\chi_{UV}$ is close to 1.

\subsection{Emergent emission.}

Once the local UV  energy density has been calculated, models are used to 
determine the C$^{+}$ and FIR emissions from each cell, assuming that they 
arise from the same area as  the one used for the calculation
of the UV radiation field. For the dust emission, we use the  model by
 D\'esert et al. (\cite {dbp}), which has  3 different components: 
PAHs which are fully ionized when $\chi_{UV}$ $>$ 1, very small grains and 
big grains. 
We calculate the dust  emission in the  four IRAS bands at 12, 25, 60 and 100
 $\mu m$ plus an additional band at 200 $\mu m$, as the reprocessing of
 the combination of the UV radiation field (912 - 2000 \AA) 
described by $\chi_{UV}$ and of the ISRF for the 2000 \AA -2 $\mu$m part of 
the spectrum. We decrease the PAH abundance in large radiation field 
environments, as suggested by Ryter et al. (\cite {ryter}), to one fifth of 
the standard value when $\chi_{UV} > 100$.

Very large molecular clouds, with masses larger than 10$^6$ M$_\odot$ 
occupy more than 4 cells in the grid. The inner cell is not directly exposed 
to the UV radiation and for this cell we assume that the ISRF 
 is attenuated by 2.5 magnitudes of visual extinction.

We assume low optical depth in the mid and far infrared. From
the analysis of the COBE maps of the Galaxy, Boulanger et al. (\cite {bp}) 
deduce  $\tau_{\lambda}$/N$_{H}$= 1 $\times$ 10$^{-25}$ cm$^{2}$ H$^{-1}$ 
($\lambda$ /250  $\mu$m)$^{-2}$, which combined with the mean column density 
of individual clouds, N$_{H}$ = 1.5 $\times$ 10  $^{22}$  cm$^{-2}$,
gives an opacity of 2 $\times$ 10$^{-3}$ at 200 $\mu m$.

According to the dominant  phase in a given cell of the model, the emergent
 infrared emission can be :\\
-- reprocessing of the whole incident stellar radiation for molecular clouds
which are  totally optically thick in the UV. All impinging radiation is 
completely reprocessed in infrared emission from the outer cells of molecular
 clouds, \\
-- proportional to the gas column density for the diffuse medium, which is
optically thin in the UV. We assume no UV extinction  at the 12 pc  scale.

We use the PhotoDissociation Region (PDR) model by Le Bourlot et al. (\cite 
{lebpr}) to estimate  the $^2P_{3/2}-^2P_{1/2}$ C$^+$ emission at the
 surfaces of molecular clouds. We use a constant molecular hydrogen density of 
5 $\times$ 10$^{3}$ H$_{2}$ cm$^{-3}$ and the incident UV field $\chi_{UV}$. 
The density is not a critical parameter as long as it is higher than the 
critical density for collisional  excitation of C$^+$ (1000 H cm$^{-3}$) 
(see Tielens \& Hollenbach \cite{th85}). Furthermore, we assume that the whole
 surface of clouds contributes to the C$^+$ emission. As for the C$^+$ 
emission from the diffuse neutral phase, we use the  model by 
Wolfire et al. (\cite {wo}) for a two phase neutral atomic medium.
Finally, the contribution from the ionized gas in the HII regions around 
the OB associations is also included. We assume that the
gas has the same density as the diffuse medium, $n_e = n_H$ and an electronic 
temperature of 10$^4$ K. The total C$^{+}$ luminosity from an HII region 
of radius $R_0$ is then proportional to the volume of the HII region  
with a correction factor to take into account the other ionization 
stages of carbon.

 Because we deal with a line, the opacity may not
be small depending on the gas distribution and viewing geometry.  
In fact, opacity effects are important at large inclination angles. To obtain
 an edge-on view of the model galaxy in the $^2P_{3/2}-^2P_{1/2}$ C$^+$ line,
 we have made an accurate calculation of the radiative transfer in
this line.  For each line of sight through the disk, we sample the line 
profile with bins of 1 kms$^{-1}$ width, and calculate the emergent intensity 
in each velocity bin, including saturation effects. We assume that the 
intrinsic velocity dispersion of a PDR is 1 kms$^{-1}$. For
the diffuse medium, the velocity dispersion is 10 kms$^{-1}$. 
We do not account for absorption in the far infrared continuum. 
We obtain $\tau_{C^+}$ = 0.40 for an edge-on view, and $\tau_{C^+}$= 0.10 
for a face-on view.

\subsection{Implementation of the model}
To avoid transient stages of the simulation, the code is evolved during a 
few time steps. We stop the simulation when  stable results are obtained
on a time scale of 20 Myrs. This time scale corresponds to about half the
 lifetime of a giant molecular cloud before disruption by photoevaporation. 
This is  the reason why we can not  integrate further in time without treating
 gas recycling. The time step has been fixed at 10$^{6}$ years, shorter than 
the lifetime of the most  massive stars. We have checked the reliability of
 the calculations by different tests: \\
-- We have verified that the total UV luminosity from the stellar population 
emerges from the galaxy either at the same wavelength, or at far infrared 
wavelengths for the light reprocessed by dust grains. The total luminosity is
conserved with an accuracy of 1 \%. \\
-- When the number of pseudo-photons leaving each OB association n$_{\gamma}$
 is too small, the map of UV radiation field is noisy  with a few extremely 
bright spots and large voids. This is due to undersampling of the galaxy 
volume. The number of pseudo-photons should be as large as possible, but we 
have verified that we obtain a good map of the UV radiation field with 
100 pseudo-photons per OB association. The map is smooth in the vicinity 
of the OB associations, hence  the ratio of FIR emissions from the diffuse 
and dense gas stays constant with increasing n$_{\gamma}$. \\
--  The cell size is also a critical parameter: since $\chi_{UV}$ is 
proportional to  $r_{cell}^{-2}$, it might be underestimated for small clouds
very close to OB associations. This has severe consequences
 for the C$^+$ emission,  which scales roughly  as 
$r_{cell}^{2} log(\chi_{UV})$,  but  little or none for the FIR emission 
which varies as   $r_{cell}^{2} \chi_{UV}$ since in that case there is no 
resultant scaling with $r_{cell}$. To test the validity of the adopted 
resolution,  we have performed a run restricted to one quadrant only, 
with a cell size of 6.1 pc. We observed no large variation  in the C$^+$ 
emission and thus conclude that the adopted resolution of  12 pc is correct 
for our purpose. Note that the volume filling factor decreases to 1.3 \% 
in the high resolution run, because we fill the space in a more accurate way 
using a higher spatial resolution.

\section{Results}
 
Table 2 summarizes the input parameters for the Standard Model, and Table 3
 presents the results. The star formation rate from 2 to 60 M$_\odot$ is 
fixed at  4.0 M$_\odot$yr$^{-1}$, with a star formation
efficiency $\epsilon$ of 5\%, as  defined in Sect. 3.4. 
With these values, the modelled H$\alpha$ luminosity 
and H$\alpha$ radial profile are in quite good agreement with the
observed data (Kennicutt \cite {kennicutt}). This is also true for the 
 UV luminosity at 2000 \AA \ and the 6 cm luminosity. We are thus confident
 that the massive star population is well constrained by the observed data. 
With a low mass cutoff at 10 M$_{\odot}$ stars, we overestimate the
 UV luminosity at 2000 \AA \ of the modelled stellar population by 20\%, 
because we miss the contribution to the UV continuum of lower mass stars, 
between 2 and 10 M$_{\odot}$.

\subsection{Disk opacity in the UV}
We have computed an average opacity over the galaxy
in the UV and for the H$\alpha$ line. We define this
opacity  as : \\
$\tau = -ln(L^{emergent}/ L^{emitted}$)
where $L^{emitted}$ is the total luminosity in the disk at a given 
wavelength and $L^{emergent}$ is the emergent luminosity. This opacity is 
computed for two different viewing angles of the model, 
$i = 0^\circ$ for face-on and $i = 90^\circ$ for edge-on. We have found a 
significant opacity for the face-on view,  at 1000 \AA, 2000 \AA \ and 
for H$\alpha$, namely  $\tau(1000 \AA)= 0.8$, $\tau(2000 \AA) = 0.7$, 
$\tau(H\alpha) = 0.60$ for the whole galaxy.

The opacity is controlled simultaneously by the geometry of the molecular
cloud ensemble and by the diffuse medium. If we ignore the extinction due to 
the diffuse component, we find an opacity of 0.51 at 1000 \AA. This value is 
due to geometrical effects, mostly blocking of the UV radiation by molecular 
clouds, and it does not depend on wavelength. Thus we can write the opacity 
at any wavelength in the UV as $\tau_{\lambda}$ = 0.51 + 
$\tau_{\lambda}^{HI}$, the second term accounting for the wavelength
dependence of the extinction in the diffuse medium. 

A global opacity of $\tau \simeq 0.8$ corresponds to a fraction of 
approximatively  45\% of the far UV stellar  radiation leaving the galaxy 
disk, mostly above or below the main  plane. Most of these photons have not 
been scattered because the probability of leaving the disc after a scattering 
event is low. This significant fraction of the radiation from massive stars
leaking out of HII regions could contribute to the maintenance of the
Reynolds layer of ionized gas. The derived face-on opacity at 2000 \AA, 0.7,
falls well within the range  of opacities derived by  Buat \& Xu (\cite {bx}).
 The  mean extinction in their sample of nearby spiral galaxies is 
$\simeq$ 0.9 mag at 2000 \AA. Though the opacity is not very large, the 
mean  distance travelled by a UV photon before absorption is quite small, 
440 pc, roughly equal to the HI disc thickness. As shown on Fig. 6, 
there are however photons travelling to much larger distances, 
1 to 2 kpc, with a small probability (0.01). Conversely,  many zones in 
the interarm receive very few UV photons. Due to the lower gas density, 
few OB associations are created in the interarm region. The numerous OB 
associations in the arms are too distant  to contribute to the local 
radiation field since the arm/interarm separation is larger than 1 kpc 
in the disk. 

The distribution of  $\chi_{UV}$ values provide further information 
on the radiation field resulting from the OB associations (Fig. 7). 
Whereas most of the galaxy is exposed to a low UV radiation field, it is 
possible to find regions with high UV intensity ($\chi_{UV} \geq 1000$) 
even at a moderate spatial resolution. The total dynamical range of the
UV radiation field extends over more than 4 orders of magnitude. This
huge variation can be explained by the close association of
OB associations and molecular clouds: in a galaxy with a prominent
spiral structure, OB associations are born in the spiral
arms, where the gas density is the highest. This maximizes both the 
illumination of molecular clouds by UV radiation and the absorption of
 UV radiation by molecular gas, hence the heating of molecular gas. 
For the model galaxy, we find that 30 \% of the total number of 
cells with molecular gas are exposed to a strong or  median radiation 
field ($\chi_{UV} \geq 10$). These cells are located in 40\% of the molecular 
clouds. This figure is comparable to the clouds in Milky Way: 
Solomon et al. (\cite {solomon}) found that in the Galaxy, at a resolution 
greater than  10 pc, 25 \% of the molecular clouds are warm and associated 
with HII  regions.  Also, Williams and Mc Kee (\cite {wmk}) estimate that 
at least one OB star is present in half of the giant molecular clouds  with
 masses larger than 10$^{5}$  M$_\odot$. The probability to find
massive stars or clusters associated with a giant molecular cloud 
increases sharply with the cloud mass and reaches almost 1 for masses 
larger than 8 $\times$ 10$^5$ M$_\odot$ (Williams \& MacKee \cite {wmk}).
Our numerical results are in agreement with these facts.

\subsection{Far infrared emission}
We now discuss the emergent radiation from the model galaxy and start with the
infrared emission. As in the D\'esert et al. (\cite {dbp}) dust model, the
luminosities in the IRAS bands  are computed as $4 \pi D^2 \nu S_{\nu}$,
where S$_{\nu}$ is the total observed flux density and $D$ is the distance to
the object. The infared colors are given as the flux density ratios, to
compare with observed data.

The model galaxy has very similar  emissions as NGC~6946 at
 60-100 \& 200 $\mu m$, with outputs of 5.1, 8.5 and 4.9 $\times$ 10$^{9}$ 
L$_{\odot}$,  corresponding to 114\%, 128\% and 144\% of the luminosities
 observed at those wavelengths. The far infrared emission comes from both 
 the molecular and atomic gas phases. 

The UV radiation is the main heating mechanism of the dense and diffuse
 gas phases, with contributions of  4.2 $\times$ 10$^{9}$ L$_{\odot}$ and 
6.1 $\times$ 10$^{9}$ L$_{\odot}$ at  60 \& 100 $\mu m$. The contribution 
to the FIR emission of the old stellar population, described by the ISRF,
 is a factor 3 lower,  with 0.9 $\times$ 10$^{9}$  L$_{\odot}$ and 
2.4 $\times$ 10$^{9}$  L$_{\odot}$ in the  60 \& 100 $\mu m$ bands. 
The situation is different at 200 $\mu m$, where dust grains heated by
the UV radiation or by the ISRF have almost equal contributions to
the total luminosity: 3.0 $\times$ 10$^{9}$ L$_{\odot}$
for the UV  and 1.9 $\times$ 10$^{9}$ L$_{\odot}$ for the ISRF.
The contribution from the inner parts of clouds illuminated by the 
attenuated ISRF  is only 0.5 $\times$ 10$^{8}$  L$_{\odot}$. As a whole,
 72 \% of the far infrared luminosity can be attributed to
UV heated gas, which is mostly molecular. The remaining 28\% 
corresponds to dust heated by the ISRF, at locations far away from the OB 
associations.

The diffuse and dense phases have similar contributions to the total 
FIR emission, with a slight excess from the molecular clouds,
 54 \% versus 46 \% from the diffuse gas. This significant contribution 
from the diffuse gas is due to the fact that it  occupies a large fraction of 
the galaxy volume.  Hippelein et al. (\cite {hippelein}) also conclude from
ISO observations of other nearby galaxies (M51, M101) that the
neutral atomic gas has an important contribution to the far 
infrared emission. The contribution from the atomic gas may be underestimated 
because we do not take into account the compression of the diffuse
gas in the spiral structure. Comparing with molecular clouds,
we can estimate that, having atomic gas concentrated in the 
spiral arms would result in a brighter FIR emission, with a slighty warmer 
color temperature since the dust grains would be closer (in average) to the
 heating sources. A precise estimate of the magnitude of the effect
is beyond the scope  of this paper. 

The global infrared excess for the model galaxy, IRE,  is defined as the 
luminosity ratio IRE= $L_{12-100 \mu m} / L_{Ly}$, with $L_{Ly}=N_{Lyc} 
h \nu_{Ly}$ and $h \nu_{Ly}$=13.6 eV. At the disk scale, the IRE  
takes the value 5.9,  in agreement with observations of Galactic HII regions 
(Caux et al. \cite {caux}, Myers et al. \cite {myers}).
 
The diffuse and dense gas (atomic and molecular) have the following 
contributions to the total luminosity of the  C$^{+}$ 158 $\mu m$ line: 
77\%  from the dense phase and 23\% from the diffuse phase. Less than 
10$^{4}$ L$_{\odot}$ comes from  HII regions.  The total emission of the 
galaxy is 2.5 $\times$ 10$^7$ L$_\odot$, a factor 2.5 lower than the 
measured value, 6.3 $\times$ 10$^7$ L$_\odot$ (Madden et al. \cite {sm}). 
Compared to the 60-100 $\mu m$ far infrared emission, the C$^{+}$ line 
represents 0.21\% of the  FIR  (60-100 $\mu m$) emission.  This figure is 
comparable to  the observed ratio  for other spiral galaxies with 
0.1 - 1 \% (Lord et al. \cite {lord}). Nevertheless, the value for NGC~6946 
was found to be  0.6 \%  (Madden et al. \cite {sm}), and 
in the Galaxy,  Shibai et al. (\cite {shibai}) and Wright et al. 
(\cite {wright})  have measured L$_{C^+}$ = 0.7 \% L$_{FIR}$ with 
the same definition of L$_{FIR}$ as above.\\

\subsection{Radial profiles}
The 60 $\mu m$ radial profile  is shown on Fig. 8a. There is a 
large decrease from the inner to the outer parts of the disk,  
about two orders of magnitude.  In NGC~6946, the same behaviour has 
been observed  by Tuffs et al. (\cite {tuffs}) using ISO. 
Averaged over the model, the S$_{60}$/S$_{100}$ infrared color appears
 to be slightly different in the two phases: 0.32 for the diffuse phase and
0.38 for the dense phase.  This FIR color decreases with
 increasing  radius from 0.40 in the center to 0.23 at R $\simeq$5 kpc
 (Fig 8b), in agreement with the maps by Engargiola (1991). 
The decrease is seen in both phases, with S$_{60}$/S$_{100}$ ranging from 
0.35 to 0.23 for the diffuse phase, and from 0.42 to 0.28 for the dense phase. 

The radial profile of the intensity of the C$^{+}$  $^2P_{3/2}$ - 
$^2P_{1/2}$ 158$\mu m$ line (Fig. 8c) shows a much flatter gradient than the 
FIR emission. This is due to the  logarithmic dependence of the line 
intensity on the incident radiation field in PDRs. As shown in Fig. 8c
and 8d, the diffuse atomic gas is the main source of C$^+$ emission at large
distance from the nucleus, for radii larger than 4 kpc. It is thus possible
to determine the intrinsic  L$_{C^+}$/L$_{FIR}$ luminosity ratio from the two 
gas phases, using the data at $R \sim 2$kpc for the molecular gas
and data at $R \geq 5$kpc for the atomic gas. We find that 
L$_{C+}$/L$_{FIR}$ is equal to 0.10 \% in the diffuse phase and 
to 0.25\% in the dense phase.

\subsection{Maps}
We show on Fig. 9 face-on maps of 100 $\mu m$,  C$^{+}$ and 
UV(912-2000 \AA) emissions. Edge-on maps at the same wavelengths are shown 
in Fig. 10 for comparison. 
 Compared to the C$^+$ observations of the edge-on galaxy NGC 891
(Madden et al. \cite {sm1}), there is an overall agreement. In particular, 
the scale height in C$^+$ is predicted
to be larger than the scale height of the CO emission, due to the contribution 
of the diffuse neutral and ionized media which have a larger scale height
(Fig. 11).

In the face-on C$^{+}$ map, there is a large hole in the interarm regions 
in the NW,  at a similar position to the hole detected by Madden et al. 
(\cite {sm}) with the KAO. This hole is due to the lower density of molecular
 gas and of OB associations in the interarm regions. Therefore few UV photons 
illuminate this region and the radiation field is very low. 
 The map shows many details and a large contrast between arm and interarm 
regions. We have smoothed the  image from the model to the resolution of the
 KAO  observations (50'' beam = 1.2 kpc at the distance of NGC~6946). 
The contrast between the brightest regions and  the
disk drops by a large factor. This resolution effect may explain the low
dynamical range found in the observed data. If PDRs are the main source of  
C$^+$ 158$\mu m$ radiation in galaxies, we predict that the emission should 
have more contrast at higher spatial resolution. This could be tested by maps 
of external galaxies made with the future Stratospheric Observatory For 
Infrared Astronomy (SOFIA).

The edge-on maps at 100 $\mu$m and in C$^{+}$ are fairly symmetrical with 
respect to the center. The edge-on C$^{+}$ map shows however a hole in the
 central region ( $r < 500 $pc) which does not appear on the
100 $\mu$m map. This hole is largely due to the large opacity for these 
lines of sight ($\tau_{C+}$ = 0.4).

\subsection{Sensitivity of the model to input parameters}
The model results are of course sensitive to the input parameters, 
therefore we have run different models deviating from the standard model by
 one parameter.

Because of the poor knowledge of the albedo in UV, we have run a model with 
a lower albedo of dust grains, $\omega$= 0.4. We find that the opacity 
increases to 1.0 at 1000 \ \AA \ \& 0.90 at 2000 \ \AA. 
The 60 $\mu m$ emission from the dense phase  increases by 5\%, while the 
100 and 200 $\mu m$ emissions  both  decrease by 10\%. This difference in 
far infrared emission is due to the moderate increase of the opacity which 
leads to a warmer dust temperature.  The effect on the emission from the 
diffuse phase is negligible.

A more extreme case is for a null albedo, suppressing any scattering effect.
 In that case, we maximize the UV opacity and the FIR emission. 
The opacity increases  to  1.25 at 1000 \AA \ and 1.01 at
 2000 \AA \  respectively.  As a result of this larger absorption, the 
60-200$\mu m$ emission  increases by 47 \% as compared to the standard model. 

In another run, we have kept the total mass of molecular gas constant, 
but used a lower mean density, 20 H$_{2}$ cm$^{-3}$ instead of 
50 H$_{2}$ cm$^{-3}$, to increase the clouds sizes. The volume filling 
factor is then 3.8 \%. These larger clouds block more light, and 30 \% only 
of the molecular cells are heated, instead of 40\% in the standard model.
As a result, the 60-200 $\mu m$ luminosity decreases by 10\%, to $16.7 
\times 10^{9}$ L$_{\odot}$.

We have also investigated the effect of the number of OB associations:
 we have kept the same star formation rate but have gathered adjacent 
associations to form more powerful sources. As a consequence,
n$_{OB}$ decreases from 12000 to 3000.  Then a smaller fraction of the 
cloud population is heated, 15\%, as compared to  40\% in the standard model. 
 But because these cells are  heated by more powerful OB associations,
 the far-infrared emission is larger and reaches $20.4 \times 10^{9}$ 
L$_{\odot}$.  Thus the FIR emission  depends slightly on the number of
 associations.  The C$^{+}$ emission decreases to $1.8 \times 10^{7}$ 
L$_{\odot}$, because of the smaller number of illuminated clouds.

If we now increase the SF efficiency, from 5 to 10\%, so as to double the
 UV luminosity,  the production rate of Lyman continuum photons increases 
by 80 \%. In that case, the mean UV opacity is 0.78. The FIR 60-200 $\mu m$ 
luminosity increases by 55 \% to 28.7 $\times$ 10$^{9}$ L$_{\odot}$.
 This shows that the FIR emission is not a linear function of the UV 
luminosity  in our model. This non-linear behaviour arises because the
 opacity is largely controlled by geometrical effects. With a larger star 
formation activity, HII regions are very large and can destroy  molecular 
clouds efficiently. Thus the mass of molecular gas decreases in the model 
with a higher SFR. This is the main reason for the non-linear behaviour.  
This result has been established with the same number of OB associations, 
while an  increased SFR will probably lead to more associations in the disk. 
However we have previously shown that the FIR emission does not depend 
strongly on the number of OB associations. 

We have investigated the effect of the atomic density on the size of 
HII regions, because we probably overestimate the  diameter of HII regions, 
using a mean atomic density and neglecting the dust absorption. 
If the local gas density is multiplied by two, the volume of
 the Str\"omgren sphere is 4 times smaller than in the standard model.  
The 60-200 $\mu m$ luminosity of dense molecular gas  increases by 
10\% to $1.1 \times 10^{10}$ L$_{\odot}$. This is explained by the reduced 
destructive effect of HII regions on molecular clouds, and then the larger 
chance for photons to be absorbed by molecular gas. The respective 
contributions from the diffuse and dense gas to the FIR(60-200 $\mu m$) 
are now  34\% and 66\%.

This last test shows that the distance between clouds and OB associations 
has a strong influence on the UV reprocessing by dense gas. For the standard 
model, we have calculated the mean distance between an OB association
and the nearest  cloud edge, d$_{OB/cloud}$, and have 
found a value of 35 pc, the mean distance between  clouds centers is
37 pc.  To have a larger separation between clouds and OB associations,
we have increased v$_{escape}$  to 30 kms$^{-1}$. 
We obtain d$_{OB/cloud}$= 39 pc. The FIR(60-200 $\mu m$) emission 
from the dense phase decreases by 15\% because of the smaller solid angles of 
the clouds viewed from the associations. As for the FIR (60-200 $\mu m$) 
from the diffuse phase, it slightly increases by 6\%. 

We have shown that part of the UV opacity is due to geometrical
effects. Indeed the UV opacity is lower when the molecular clouds
are distributed uniformly in the disk. We have used an earlier epoch of the
simulation, when the distribution of gas clouds is axisymmetric. We have
kept the same value for the other  parameters  (number of OB associations, 
star formation rate, etc.). In that case, the clouds occupy a larger fraction
volume of the galactic disk, and the mean distance between clouds increases.
Because of this larger mean distance
between clouds,  the  opacity at 1000 \AA \ decreases to 0.47.

\section{Discussion \& conclusions}

We have shown that with simple assumptions about the birth of massive stars
 and their relationships with the ISM, we can reproduce qualitatively and 
quantitatively the  characteristics of the UV, H$\alpha$ and FIR emissions of
 a particular object,the Sc galaxy NGC~6946.  For such a galaxy with a 
prominent spiral structure, having a large mass of neutral gas, and 
forming stars actively, the observed far infrared emission is produced both 
in molecular gas and in the diffuse atomic gas. More precisely, 
54 \% of the FIR (60-200 $\mu m$) emission comes from dust grains in 
giant molecular clouds. Dust in the  diffuse neutral atomic gas
 contributes to  about 46 \% of the total FIR luminosity.

 We have evaluated the respective contributions of the UV radiation from 
massive stars and of the radiation field from  the old stellar population. 
We find that 72 \% of the FIR luminosity can be attributed to UV heated 
dust grains, which reside mostly in molecular clouds envelopes.  
The remaining 28 \% is due to dust heated by the radiation field from 
the old stellar population at locations far away from OB associations.

We have calculated the emission of the model galaxy in the 
$^2P_{3/2}-^2P_{1/2}$ fine structure line of C$^+$ at 158 $\mu m$. 
In the spiral arms, photon dissociation regions at 
the surfaces of molecular clouds are the main source of the emission.
We have found a large arm-interarm contrast in this line. This effect
could be tested by high angular resolution maps of galaxies. It
 results naturally from the combination of a lower gas density and lower 
radiation field in the interarm regions, because of the short mean free 
path for UV photons, $\sim$ 440 pc. As a whole, PDRs represent 76\% of 
the emission.  The contribution from the diffuse phase is found 
to be $\sim$ 24 \%. Our model is able to account for about 40 \% of the 
observed  C$^+$ emission of NGC~6946.  The emission from PDRs should be viewed 
as a lower limit since we use the model by Le Bourlot et al. (\cite {lebpr})
 with low abundances of carbon and other elements in the gas phase:
 [C]/[H]= 3 $\times$ 10$^{-5}$. The  average value is  1.3$\times$ 10$^{-4}$ 
for Galactic diffuse clouds (Snow \& Witt \cite {sw}), a factor  of 4  larger 
than the value used in the model. Since the C$^+$ 158 $\mu m$ emission 
scales roughly with the column density,  hence the carbon abundance, the 
total C$^+$ luminosity from PDRs could be larger by at least a factor 
three than our current model prediction. This would increase the contribution 
from PDRs to the total C$^+$ emission of the model galaxy: with this scaling 
factor, the predicted C$^{+}$ luminosity of PDRs would reach  
$6 \times 10^{7} L_{\odot}$. Moreover, the C$^{+}$ emission from the 
diffuse gas is overestimated, because the model of Wolfire et al. (\cite {wo})
 assumes [C]/[H] $\sim$ 3 $\times$ 10$^{-4}$ in the gas. 
Thus the diffuse emission could be 2-3 times smaller than in our 
standard model. Adopting [C]/[H]=1.3 $\times$ 10$^{-4}$ 
in both phases would thus enhance the differences of $L_{C^{+}}/L_{FIR}$ 
between dense and diffuse gas.

The knowledge of the cloudy nature of the ISM, and of the global structure 
of the galaxy, is important to determine  how far UV photons can travel 
away from OB associations. The filling factor and the mass/radius scaling 
law appear to be  major parameters for the transfer of stellar radiation 
in the galaxy disk, because they determine at the same time the obscuration 
and the size of the emitting regions. Other important parameters are the 
number of OB associations and the sizes of HII regions,  because with a 
large number of OB associations or with small HII regions, molecular clouds 
are on average closer to massive stars, and are thus more efficiently heated.  

In all the models we ran, we have found that the
average internal UV opacity is of the order 0.8. The discs
are therefore  moderately opaque in the UV, as measured by Buat \& Xu 
(\cite {bx}). This moderate opacity holds for face-on discs. 
Edge-on discs are quite opaque, with a small fraction of the luminosity 
escaping, less than  1.0\% of the face-on  luminosity.  This fraction
 corresponds to an equivalent extinction of 5 magnitudes in the UV.

These results have been obtained using a crude description
of the interstellar  medium. The adopted spatial resolution
results from a compromise between astrophysical requirements 
and computational needs, but is certainly very poor compared to the complexity
of the interstellar medium. The good agreement of the observed
and predicted large scale properties shows nevertheless that the transfer 
of UV radiation, and the role of the radiation for the gas and dust heating, 
are correctly described at the 12 pc scale.
This is in agreement with previous works estimating that dust heating 
by UV radiation occurs principally  at large distances from massive stars  
(Murthy et al. \cite {murthy}, Leisawitz \& Hauser \cite {lh}).

\begin{acknowledgements}
We have benefited from the help of F. Viallefond for the numerical 
calculations and the data processing, and from discussions with D. Beck, 
G. Helou, S. Madden,  S. Shore. We thank F.X. D\'esert and
 J. Le Bourlot for letting us use their codes, and
M.G. Wolfire for providing  unpublished results. 

\end{acknowledgements}|

\newpage

\begin{table*}
\caption{Observed parameters for NGC 6946.}
\begin{tabular}{|c|c|c|}
\hline
\multicolumn{1}{|c|}{Observed data} &
\multicolumn{1}{|c|}{Value}  &
\multicolumn{1}{|c|}{Reference} \\ 
 \hline
Distance   & 5 Mpc &  De Vaucouleurs (1979)\\
Inclination  & 34$^\circ$ & Consid\`ere \& Athanassoula (1988)\\
 M$_{H2}$ (R $<$ 6') & 2.0 10$^{9}$ M$_{\odot}$ $^{1}$&  Young \& Scoville 
(1982)\\
 M$_{HI}$ (R $<$ 6') & 2.6 10$^{9}$ M$_{\odot}$ $^2$ & Boulanger \& Viallefond 
(1992)\\
L$_{1950-2050 A}$ & 5.4 10$^{8}$  L$_{\odot}$ $^3$ & Buat et al. (\cite 
{bdh})\\
L$_{H \alpha}$ &1.0 10$^{8}$ L$_{\odot}$ $^4$ & DeGioa et al. (\cite 
{degioa}))\\
B$_{t}^{0}$ &8.49 mag& RC3\\
IRAS 12$\mu$m  & 2.3 10$^{9}$ L$_{\odot}$ $\pm$ 20 \% $^5$ & Engargiola (1991)\\
IRAS 25$\mu$m  & 1.4 10$^{9}$ L$_{\odot}$ $\pm$ 20 \% $^5$ & Engargiola (1991)\\
IRAS 60$\mu$m  & 4.5 10$^{9}$ L$_{\odot}$ $\pm$ 20 \% $^5$  & Engargiola 
(1991)\\
IRAS 100$\mu$m & 6.6 10$^{9}$ L$_{\odot}$ $\pm$ 20 \% $^5$ & Engargiola (1991)\\
IRAS 200$\mu$m & 3.4 10$^{9}$ L$_{\odot}$ $\pm$ 20 \% $^5$ & Engargiola (1991)\\
L$_{C^{+}}$ &6.3 10$^{7}$ L$_{\odot}$ $^6$ & Madden et al. (\cite {sm})\\
S$_{6cm}^{thermal}$ & 83 mJy $\pm$ 25 $^7$  & Klein et al. (\cite {klein})\\
 \hline
\end{tabular}
\\
We use the value 3.8 $\times$ 10$^{33}$ erg s$^{-1}$ for the solar 
luminosity at any wavelength.\\

$^1$ We use N$_{H_2}$/I$_{CO(1-0)}$ = 2.3 10$^{20}$ mol 
cm$^{-2}$/(K kms$^{-1}$) and a total intensity 
I$_{CO(1-0)}$= 569 Kkms$^{-1}$ in a 45'' beam, we 
do not account for projection effects.\\
$^2$ We use M$_{HI}$ = 1.9 $\times$ 10$^{10}$ M$_{\odot}$ at a distance of 
10 Mpc, corresponding to $4.7 \times 10^9$ M$_{\odot}$ for a distance of 5 Mpc.
By performing an integration over the HI map, we estimate that the  part 
with R $<$ 6' contributes to  55 \% of the global emission.\\
$^3$ The 2000 \AA \ flux,  6.92 $\times$ 10$^{-12}$ erg 
cm$^{-2}$s$^{-1}$\AA$^{-1}$,  is corrected for the Galactic extinction and 
integrated over a 100 \AA \ band. Note that Donas \& Deharveng (\cite {donas})
have reported a flux at 2000\ \AA \ twenty times lower
(1.66 $\times$ 10$^{-13}$ erg cm$^{-2}$s$^{-1}$\AA$^{-1}$). Measurements 
at other UV  wavelengths are available in Rifatto et al. (1995).\\
$^4$ We use the uncorrected flux f$_{H\alpha}$= 
1.31 $\times$ 10$^{-10}$ erg cm$^{-2}$s$^{-1}$ and the relationship 
L$_{H\alpha}$ = 3.13 $\times$ 10$^{16}$ D$_{Mpc}^{2}$ f$_{H\alpha}$
from Young et al. (\cite {young}). We have subtracted the contribution
from the nucleus, estimated to be 20 \% of the total luminosity. Note that 
Young et al. (\cite {young}) give a lower total flux, $3.38 \times 10^{-11}$
 erg cm$^{-2}$s$^{-1}$, while  Kennicutt (\cite {kennicutt}) reports 
 a mean surface brightness of $4.2 \times 10^{32}$ erg s$^{-1}$ pc$^{-2}$ 
corresponding to  a total H$\alpha$ luminosity of $3 \times 10^{7}$ 
L$_{\odot}$ at the adopted distance.\\ 
$^5$ We estimate the luminosity in the band centered on frequency
$\nu$, by $\nu F_{\nu}$. We use the IRAS and KAO fluxes reported by Engargiola 
(\cite {engargiola}) for a radius R $<$ 5.6' and 
correct from the contribution of the nucleus (central 45'') to obtain 
12 Jy at 12$\mu$m, 15 Jy at 25$\mu$m, 114 Jy at 60$\mu$m, 283 Jy at 100$\mu$m 
and 288 Jy at 200$\mu$m. The global color for the disk is 
S$_{60}$/S$_{100}$ = 0.40. The far infrared luminosity is 
L$_{FIR (\ 60-100 \mu m)}$= L$_{60 \mu}$ + L$_{100 \mu}$ = $1.1 \times
 10^{10}$ L$_{\odot}$, L$_{FIR (\ 60-200 \mu m)}$ = $1.45 \times 10^{10}$ 
L$_{\odot}$\\
$^6$ We have excluded the contribution of the nucleus and 
rescaled the data by Madden et al.  (\cite {sm}) for the adopted distance. \\
$^7$ We use an estimation of 625 mJy for the total flux at 6cm (Klein et al. 
\cite {klein}),  subtract the nuclear contribution estimated to be 33\% of
the total flux, and we  keep 20\% of the resultant disk flux as 
the thermal component.\\
\end{table*}

\begin{table*}
\caption{Input parameters for the standard model.}
\begin{tabular}{|c|c|}
\hline
\multicolumn{1}{|c|}{Parameter} &
\multicolumn{1}{|c|}{Adopted value}  \\
 \hline
cell resolution &12.2 pc   \\
HI half thickness & 180 pc   \\ 
H$_{2}$ scaleheight & 65 pc  \\
mean V$_{OB}$ & 10 kms$^{-1}$  \\ 
n$_{\gamma}$ & 100   \\ 
time step & 10$^{6}$ years   \\ 
number of molecular clouds & 18000 \\
number of OB associations & 12000  \\
star formation rate (2 -- 60 M$_{\odot}$) & 4.0 M$_{\odot}$ yr$^{-1}$  \\
volume filling factor of H$_{2}$  &  2.5 \%  \\
\hline
\end{tabular}
\end{table*}

\pagebreak
\newpage

\begin{table*}
\caption{Output values from  the simulation for the standard model.}
\begin{tabular}{|c|c|c|c|c|c|}
\hline
\hline
\multicolumn{1}{|c|}{Parameter} &
\multicolumn{1}{|c|}{Total} &
\multicolumn{1}{|c|}{Dense} &
\multicolumn{1}{|c|}{Diffuse} &
\multicolumn{1}{|c|}{UV} &
\multicolumn{1}{|c|}{ISRF} \\
\hline
L$_{912- 2000 A}$  (L$_{\odot})$ & 1.3 10$^{10}$ &--- &--- &--- &--- \\
L$_{1950- 2050 A}$  (L$_{\odot})$ & 5.2 10$^{8}$ &--- &--- &--- &--- \\
L$_{H \alpha}$ (L$_{\odot}$) &1.2 10$^{8}$  &--- &--- &--- &---\\
B$_{t}^{0}$  (mag)  & 8.96  &--- &--- &--- &--- \\
L$_{12}$  (L$_{\odot}$) &2.1 10$^{9}$ &1.0 10$^{9}$ &1.0 10$^{9}$ &1.7 
10$^{9}$ &0.3 10$^{9}$ \\
L$_{25}$  (L$_{\odot}$)           &2.0 10$^{9}$ &1.0 10$^{9}$ &1.0 10$^{9}$ 
&1.7 10$^{9}$ &0.3 10$^{9}$ \\
L$_{60}$  (L$_{\odot}$)           &5.1 10$^{9}$ &2.9 10$^{9}$ &2.2 10$^{9}$ 
&4.2 10$^{9}$ &0.9 10$^{9}$ \\
L$_{100}$  (L$_{\odot}$)          &8.5 10$^{9}$ &4.5 10$^{9}$ &4.0 10$^{9}$ 
&6.1 10$^{9}$ &2.4 10$^{9}$ \\
L$_{200}$  (L$_{\odot}$)          &4.9 10$^{9}$ &2.5 10$^{9}$ &2.5 10$^{9}$ 
&3.0 10$^{9}$ &1.9 10$^{9}$ \\
L$_{60 - 200}$  (L$_{\odot}$)          &18.6 10$^{9}$ &9.9 10$^{9}$ &8.6 
10$^{9}$ &13.3 10$^{9}$ &5.2 10$^{9}$ \\
L$_{C +}$ (L$_{\odot}$)           &2.6 10$^{7}$ &2.0 10$^{7}$ &0.6 10$^{7}$ 
&--- &--- \\
S$_{6cm}$ (mJy)& 124  &--- &--- &--- &--- \\ 
L$_{C^{+}}$/L$_{FIR \ (60-100)}$ & 0.18 \% & 0.25\%& 0.10 \% &--- &---\\
 \hline
\end{tabular}

All luminosities are emergent luminosities. The total UV luminosity generated 
by \\
 OB associations in the disk between 912 and 2000 \AA \ is $3.10 \times
 10^{10}$ L$_{\odot}$.

\end{table*}

\newpage

{\bf Figure captions}
\\

\noindent
Fig. 1.  CO(J=2--1) map at 13'' resolution  obtained with the IRAM 30m
 radiotelescope. \\
Fig. 2. V-I image, obtained with the 1.20m telescope at
the  Observatoire de Haute Provence (P. Boiss\'e, private communication). 
The field of view is 10' by 10', the pixel size is 2.3''. The gray scale 
runs from white for  blue colors to black is for  red colors.\\
Fig. 3a.  R band image of NGC 6946 used as input for
the calculation of the gravitational potential. Field stars have been removed,
the image has then been rotated and deprojected  to get a face-on view of 
NGC~6946. \\
Fig. 3b. A face-on view of the  model galaxy, with molecular clouds drawn as
circles  and OB associations drawn as stars.  Only 20 \% of the OB 
associations and molecular clouds are drawn.\\
Fig. 3c. A close-up view of Fig. 3b.  The clouds are drawn at their exact
size in the model.\\
Fig. 4. Adopted rotation curve (dot-dashed line), angular frequency 
$\Omega (r)$ (full line), and $\Omega \pm \frac{\kappa}{2}$ curves.
The corotation for the adopted pattern speed is located at $r = 3.5$ kpc.\\
Fig. 5. Distribution of the production rate of Lyman continuum photons 
$N_{Lyc}$,  for the modelled population of  OB associations (full line) 
in NGC~6946, and for the Galaxy  (dot-dashed curve) (McKee \& Williams (1997)).
The plot can be read for instance as  500 OB associations out of 
12000 have N$_{Lyc}$ $>$ 10$^{50}$ s$^{-1}$.\\
Fig. 6. Distribution of the distance from their parent OB association, 
travelled by UV photons  before absorption. The median distance is 120 pc,
 and the mean distance is 440 pc.\\
Fig. 7. Distribution of the UV intensity  measured relative to the ISRF
in the modelled galaxy, for cells filled with diffuse gas (dot-dashed line)
and for cells filled with molecular gas (full line).\\
Fig. 8a. Radial profiles of the 60 $\mu m$ surface brightness. The thin 
solid line shows the combination of the two gas phases, while the triangles 
show the contribution of the diffuse atomic gas, and the squares the 
contribution from the molecular gas. The thick solid line  shows the mean  
radial profile deduced by Engargiola (1991) from an IRAS image.\\
Fig. 8b. Radial profiles for the ratio of 60 and 100 $\mu m$ fluxes, 
global and for the two phases.\\
Fig. 8c. Radial profiles of the C$^{+}$ emission: global and for 
the two phases.\\
Fig. 8d. Radial profiles of L$_C^{+}$/L$_{(60-100 \mu m)}$: global and 
for the two phases.\\
Fig. 9. Face-on views of:\\
a) 100 $\mu m$ emission, at 48 pc resolution. The gray scale ranges 
from 1 to 10$^3$ L$_{\odot}$ pc$^{-2}$. We have overlaid 
 contours of the same map convolved with a 750 pc beam: 
the levels are at 10, 30, 60, 100, 200 L$_{\odot}$ pc$^{-2}$. \\
b) C$^{+}$ line, at  48 pc resolution. The gray scale ranges from 
10$^{-2}$ to 1 L$_{\odot}$ pc$^{-2}$. Overlaid
 contour levels from 0.1 to 0.6 by 0.1 L$_{\odot}$ pc$^{-2}$ for the
same image convolved with a 750 pc beam. \\
c) emergent UV surface brightness (resolution: 48 pc). 
The gray scale ranges from 10$^{-2}$ to 10$^5$ L$_{\odot}$ pc$^{-2}$. \\
Fig. 10. Edge-on views of NGC 6946, at  48 pc resolution. The
 linear scale is not identical for both axes. \\
a) 100 $\mu m$ IRAS band, with contour levels at 
10$^{3}$,  $3 \times 10^{3}$,  $6 \times 10^{3}$ and 10$^{4}$ 
L$_{\odot}$ pc$^{-2}$.\\
b)  C$^{+}$, at the same resolution, accounting for the line opacity. 
The gray scale ranges from 0 to 50 L$_{\odot}$ pc$^{-2}$. 
Contour levels from 10 to 50 by 10 L$_{\odot}$ pc$^{-2}$.\\
Fig. 11. Average vertical profiles through the disk  for the C$^+$ 
(dot-dashed line)  and CO(1-0)(full line) emissions. The
CO(1-0) profile  presents a smaller scaleheight (40 pc) than the 
C$^{+}$ profile (100 pc). We have assumed that the CO(1-0) emission is
 proportional to the molecular gas column density.\\

\end{document}